\documentclass[11pt,aps,superscriptaddress,preprintnumbers,nofootinbib]{revtex4-1}
\pdfoutput=1

\usepackage{amssymb,amsmath}
\newcommand{\BigO}[1]{\ensuremath{\operatorname{O}\left(#1\right)}}

\usepackage{graphicx}
\usepackage{color}
\usepackage{float}
\usepackage{bbold}

\usepackage{hyperref}
\usepackage{cleveref}

\def\lesssim{\mathrel{\mathpalette\vereq<}}
\def\gtrsim{\mathrel{\mathpalette\vereq>}}

\begin{document}
	
\preprint{UTTG-19-15}
\preprint{TCC-008-15}


\title{Can A Pseudo-Nambu-Goldstone Higgs Lead To Symmetry Non-Restoration?}

\author{Can Kilic}
\author{Sivaramakrishnan Swaminathan}
\affiliation{Theory Group, Department of Physics and Texas Cosmology Center,
\\The University of Texas at Austin,  Austin, TX 78712 U.S.A.}


\begin{abstract}

The calculation of finite temperature contributions to the scalar potential in a quantum field theory is similar to the calculation of loop corrections at zero temperature. In natural extensions of the Standard Model where loop corrections to the Higgs potential cancel between Standard Model degrees of freedom and their symmetry partners, it is interesting to contemplate whether finite temperature corrections also cancel, raising the question of whether a broken phase of electroweak symmetry may persist at high temperature. It is well known that this does not happen in supersymmetric theories because the thermal contributions of bosons and fermions do not cancel each other. However, for theories with same spin partners, the answer is less obvious. Using the Twin Higgs model as a benchmark, we show that although thermal corrections do cancel at the level of quadratic divergences, subleading corrections still drive the system to a restored phase. We further argue that our conclusions generalize to other well-known extensions of the Standard Model where the Higgs is rendered natural by being the pseudo-Nambu-Goldstone mode of an approximate global symmetry.

\end{abstract}

\maketitle


\section{Introduction}

Naturalness of the Standard Model (SM) requires the cancellation of divergent contributions to the Higgs mass at the loop level. Most known solutions of the little hierarchy problem involve introducing new particles that cancel the divergences caused by their SM partners, where the cancellation relies on the existence of a symmetry. In the case of supersymmetry (SUSY) the symmetry in question is a spacetime symmetry that relates bosons and fermions whereas models that realize the Higgs field as a pseudo-Nambu-Goldstone boson (pNGB) accomplish this with an internal symmetry. To be precise, in this paper we will assign a very specific meaning to the word ``natural'', namely we will label a model as natural if (after cancellations) any existing quadratically divergent contributions to the Higgs potential are of the same order as, or negligible to the leading logarithmic contributions.

In the SM, electroweak symmetry is restored at temperatures above ${\mathcal O}(100)$~GeV\cite{Kirzhnits:1972ut,Kirzhnits:1974as,Kirzhnits:1976ts,Carrington:1991hz,Arnold:1992rz,Espinosa:1992kf}. Extensions of the SM display a similar behavior at finite temperature. In particular, finite temperature breaks SUSY, and therefore the diagrams whose quadratic divergences cancel each other at zero temperature no longer cancel at finite temperature, generating a thermal mass for the Higgs proportional to $T^{2}$ and restoring electroweak symmetry. On the other hand, there is no a priori reason why the cancellation of quadratic divergences should not persist at finite temperature for models with same-spin partners. It was investigated in ref.~\cite{Espinosa:2004pn,Aziz:2009hk,Ahriche:2010kh} whether this may lead to the existence of a broken phase of electroweak symmetry at high temperature in Little Higgs models~\cite{ArkaniHamed:2001ca,ArkaniHamed:2001nc,ArkaniHamed:2002pa,ArkaniHamed:2002qx,ArkaniHamed:2002qy}\footnote{Of course, the subject of possible symmetry non-restoration has a long history that significantly predates the Little Higgs mechanism, starting with refs.~\cite{Weinberg:1974hy,Mohapatra:1979qt,Mohapatra:1979vr}.}. In the following, we will perform a similar analysis on a theory with same-spin partners.

A calculation keeping only the quadratically divergent but not subleading contributions to the Higgs finite temperature effective potential can be justified when SM partners thermally populate the plasma, which happens only close to the cutoff of the effective field theory. However, in theories such as the Littlest Higgs~\cite{ArkaniHamed:2002qy} where the Higgs is nonlinearly realized as a pNGB, higher dimensional terms in the effective potential can become important close to the cutoff due to power-law divergent contributions. Then, a one-loop analysis may prove insufficient for calculations at energies above the decay constant $f$ of the sigma model. In fact, the effective field theory of the Littlest Higgs nonlinear sigma model becomes strongly coupled well below $4 \pi f$ due to higher-dimensional operators being corrected by scalar loops~\cite{Espinosa:2006nu}. Thus, the analysis in ref.~\cite{Espinosa:2004pn} with the Littlest Higgs EFT is untrustworthy for $T \gtrsim f$ since Matsubara modes have masses of order $\pi T$. For this reason, we will choose a benchmark model in this paper which has a weakly coupled linear UV completion, namely the Twin Higgs \cite{Chacko:2005pe}, where a one-loop calculation should be reliable.

In our calculation, we will include subleading corrections in the finite temperature potential, which are of a size comparable to the zero-temperature effective potential, and therefore cannot be neglected. We find that while we agree with ref.~\cite{Espinosa:2004pn} that the thermal corrections of $\BigO{T^2}$ do cancel, subleading corrections still restore the symmetry at high temperature in the Twin Higgs model. Furthermore, we will argue that our conclusions extend beyond the Twin Higgs model, and should remain valid in models where the cancellation of $\BigO{T^2}$ corrections to the Higgs potential are ensured by an approximate global symmetry of which the Higgs is a pNGB, and therefore electroweak symmetry is generically restored at high temperature in models that are natural according to our definition.

This paper is organized as follows: We review the salient features of the Twin Higgs model in \cref{sec:THmodel}, followed by a review of the general aspects of calculating the finite temperature effective potential in \cref{sec:FTgen}. We then calculate the finite temperature effective potential for our benchmark model and we present the results in \cref{sec:THfinitetemp}. In \cref{sec:conc} we consider the symmetry structure of other well-known natural extensions of the SM where the Higgs is realized as a pNGB and we conclude that the lessons learned from the benchmark model are generic.


\section{The Twin Higgs model}
\label{sec:THmodel}

There are several variations keeping with the spirit of the Twin Higgs setup
\cite{Barbieri:2005ri,Chacko:2005pe,Chacko:2005vw,Chacko:2005un,
Foot:2006ru,Goh:2007dh,Craig:2014aea,Geller:2014kta,Barbieri:2015lqa,Low:2015nqa,Craig:2015pha}
and here we adopt a minimal version of the model presented in ref. \cite{Craig:2015pha} as a benchmark model, and limit ourselves to a description of the aspects most relevant to our purposes. The reader is invited to consult the original references for any additional details not presented here.

In very rough terms, Twin Higgs models introduce a second set of degrees of freedom identical to the SM.
The second set of fermion fields fill out the same gauge representations under the new gauge groups as the SM fermions do under the SM gauge groups.
The two sectors couple to each other through the scalars (Higgs), and in our benchmark model, they are both charged under $U(1)_{Y}$.
Furthermore, an approximate $\mathbb{Z}_{2}$ symmetry relates these two sectors (with sector A being identified as the SM). Since the main interest in constructing the Twin Higgs setup is to keep contributions to the Higgs potential under control, in many phenomenological studies, all fermion fields are neglected, for simplicity, except those that are relevant for cancelling the divergent contributions due to the top Yukawa coupling, and this is the approach that we adopt as well.

For the purposes of this study, we will take the gauge symmetry of the theory to be
\begin{equation}
G=\left[SU(3)\times SU(2)\right]^{2}\times U(1)_{Y}\equiv \left[SU(3)\times SU(2)\right]_{A}\times \left[SU(3)\times SU(2)\right]_{B}\times U(1)_{Y},
\end{equation}
and the relevant fermionic degrees of freedom in the top sector fill out the representations $Q_{A,B}=\left(3,2\right)_{A,B}$, $T^{c}_{A,B}=\left(\bar{3},1\right)_{A,B}$ with hypercharges $1/6$ and $-2/3$, respectively. Under the $\mathbb{Z}_{2}$ symmetry, the gauge and matter fields of the $A$ and $B$ sectors are exchanged (and the $U(1)_{Y}$ is unaffected).

It should be noted that this choice of the gauge sector is not phenomenologically viable. In particular since there is only one $U(1)$ factor, the heavy $Z'$ particle inherits couplings to the SM fermions that are experimentally excluded. Adding a second $U(1)$ factor without additional model building in the exact $\mathbb{Z}_{2}$ limit is also problematic, since it leads to the existence of a second massless photon. A number of phenomenological studies of the Twin Higgs model and its variants have focused on these and other issues~\cite{Goh:2006wj,Jung:2007ea,Abada:2007sm,Dolle:2007ce,
Batra:2008jy,Liu:2008ca,Liu:2008bx,Liu:2008zzb,
Liu:2009ak,Ma:2009kj,Liu:2009zze,Goh:2009fm,
Shen:2010zzh,Liu:2010jp,Liu:2010nn,
Shen:2011zz,Shen:2011zzg,Zhang:2011as,Wang:2011zzt,Ma:2011zzb,Wang:2011zza,Liu:2011zzi,Wang:2011mra,
Zhan-Ying:2013kxa,Liu:2013dma,Liu:2013gpy,
Han:2014fla,Liu:2014pts,Liu:2014rqa,Burdman:2014zta,Craig:2014roa,Liu:2014uua,
Hetzel:2015bla,Schwaller:2015tja,Han:2015orc,Garcia:2015loa,Craig:2015xla,Garcia:2015toa,Farina:2015uea,Curtin:2015fna}
however for the purposes of this paper we choose to work with this very minimal model. While extended models exist that address such phenomenological issues, using such a model would only obscure the simple idea behind our analysis without significantly altering our conclusions
\footnote{Most extended models need to introduce additional breaking of the $\mathbb{Z}_{2}$ symmetry, and deviations from the exact symmetry limit tend to reintroduce quadratic divergences which lead to $\BigO{T^2}$ symmetry restoring mass terms.}.

The cancellations to the Higgs mass arise from an approximate $SU(4)$ global symmetry in the scalar sector, of which the $SU(2)_{A}\times SU(2)_{B}$ subgroup is gauged. The scalar degrees of freedom belong to the fundamental representation of this global $SU(4)$ symmetry, such that under the gauged subgroup they transform as
\begin{equation}
	H \equiv \begin{bmatrix} H_A \\ H_B \end{bmatrix} \longrightarrow \left[{
    \begin{array}{c|c} SU(2)_A &  \\ \hline & SU(2)_B \end{array}}\right] \begin{bmatrix} H_A \\ H_B \end{bmatrix}.
\end{equation}
Up to a term that will be added later, the tree level potential for the scalars is chosen to respect the global $SU(4)$ symmetry,
\begin{equation}
V(H) = \frac{\lambda}{4} {\left( {|H|}^2 - {f}^2 \right)}^2.
\label{eq:scalarpot}
\end{equation}
The $SU(4)$ symmetry is spontaneously broken down to $SU(3)$ as $H$ acquires a vacuum expectation value (VEV), which results in seven Nambu-Goldstone bosons and a heavy radial mode. Below the scale $f$, the radial mode can be integrated out to obtain a nonlinear sigma model for the degrees of freedom parameterized as
\begin{equation}
\exp \dfrac{i}{f} \left[{
\begin{array}{ccc|c} & & & h_1 \\
& \text{\huge 0} & & h_2 \\ & & & h_3 \\
\hline h_1^\ast & h_2^\ast & h_3^\ast & h_0
\end{array}}\right]
{\begin{bmatrix} 0 \\ 0 \\ 0 \\ f
\end{bmatrix}}
\equiv
\begin{bmatrix}
i f \frac{h}{\sqrt{h^\dagger h}} \sin \left(\frac{\sqrt{h^\dagger h}}{f}\right) \\i f \frac{h'}{\sqrt{{h'}^\dagger h'}}  \cos \left(\frac{\sqrt{h^\dagger h}}{f}\right)
\end{bmatrix},
\end{equation}
which defines $h$ as the SM Higgs doublet field, and $h'$ as the twin Higgs which is charged under the twin $SU(2)$. It is straightforward to see that to leading order, $h=H_{A}$.

The global $SU(4)$ is broken down to ${SU(2)}_A \times {SU(2)}_B$ when the theory is gauged, and once the Yukawa interactions are introduced, where $H_{A}$ couples $Q_{A}$ and $T^{c}_{A}$, and $H_{B}$ couples $Q_{B}$ and $T^{c}_{B}$.
\begin{equation}
	{\mathcal L}_{\textrm{Yukawa}} = y \left(H_A^\dagger Q_A T_A^c +  H_B^\dagger Q_B T_B^c\right)
\end{equation}
Note that these terms are compatible with the $\mathbb{Z}_2$ symmetry even though they explicitly break the $SU(4)$. This has a very important consequence: one-loop corrections to the quadratic part of the scalar potential respect the $\mathbb{Z}_{2}$,
that is,
they are proportional to $H_A^\dagger H_A + H_B^\dagger H_B$, which can be written as $H^\dagger H$
\footnote{In fact, quadratically divergent mass corrections have this property to all loop orders. Even if the $Z_2$ symmetry is softly broken by the $\mu^2$ term to be introduced later in this section, there will be higher-loop mass corrections proportional to $\mu^2$ but those are not quadratically divergent.}.
In other words, leading quantum corrections to the quadratic part of the potential accidentally respect the full global $SU(4)$ symmetry. Specifically, corrections from the Yukawas and the $SU(2)$ gauge groups have the following form:
\begin{equation}
V_1(H) \supset \left[ - \frac{3 y^2 \Lambda^2}{8 \pi^2} + \frac{9 g^2 \Lambda^2}{64 \pi^2} \right] \left( H_A^\dagger H_A + H_B^\dagger H_B \right).
\end{equation}
\begin{figure}
    \begin{center}
		\includegraphics[scale=1]{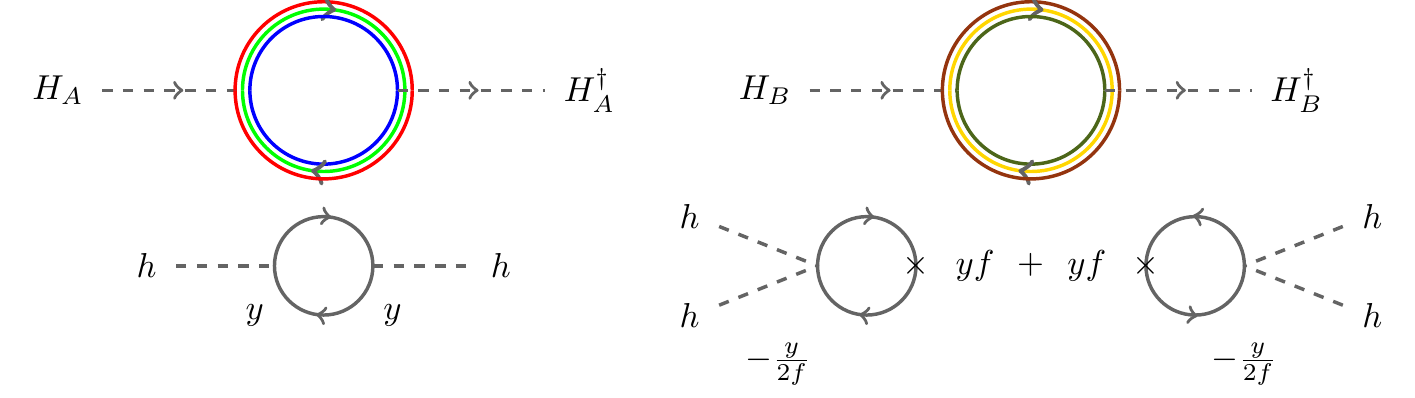}
    \end{center}
    \caption{The cancellation of quadratic divergences in the top sector in terms of Feynman diagrams in the linear (top row) and non-linear (bottom row) formulations of the model.}
    \label{fig:cancellations}
\end{figure}
Therefore, any quadratically divergent contributions give a mass to the radial mode of $H^{\dagger} H$, but not the SM Higgs doublet $h$. This cancellation is easiest to see from the linear theory, and appears to be somewhat mysterious from the point of view of the low energy theory due to an unusual four-point coupling between the SM Higgs doublet and the partner fermions. This is illustrated in figure~\ref{fig:cancellations}. The $\mathbb{Z}_2$ similarly prevents divergent contributions from the gauge sector, which again is most easily seen in the linear theory, but of course this also holds true in the nonlinear sigma model of the low energy theory after the radial mode has been integrated out.

More quantitatively, the one-loop Coleman-Weinberg potential (in Landau gauge)
\begin{equation}
V_{CW}(H) = \frac{1}{64 \pi^2}{\rm STr}\left[ m^4 (H) \left( \log \left( \frac{m^2 (H) }{\Lambda^2} \right) - \frac{3}{2} \right) \right] \label{eq:CW}
\end{equation}
includes contributions from the top sector, where
\begin{equation}
m^2_{t_A} = y^2 f^2 \sin^2 \left( \frac{v}{\sqrt{2} f} \right)
\ \ {\rm and }\ \ m^2_{t_B} = y^2 f^2 \cos^2 \left( \frac{v}{\sqrt{2} f} \right)
,
\end{equation}
with $\langle h\rangle=\frac{1}{\sqrt{2}}\left(v,0\right)$, and from the gauge sector, where, in the $g_{U(1)_{Y}}\rightarrow0$ limit
\begin{equation}
m^2_{W_A} = \frac{g^2 f^2}{2} \sin^2 \left(\frac{v}{\sqrt{2}f}\right) \equiv \frac{g^2 v^2_{\mathrm{EW}}}{4}
\ \ {\rm and }\ \ m^2_{W_B} = \frac{g^2 f^2}{2} \cos^2 \left(\frac{v}{\sqrt{2}f}\right)
\label{eq:gbmasses}
\end{equation}
Since the tree level potential thus far respects the $SU(4)$ symmetry, no potential for $h$ is generated from the scalar sector at one loop. Thus $h$ only acquires a mass at one-loop through the top and gauge sectors, with the former dominating over the latter.

\begin{figure}
    \begin{center}
		\includegraphics[scale=1]{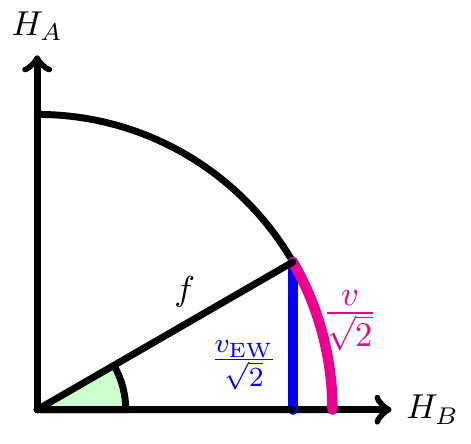}
    \end{center}
    \caption{Graphical illustration of parameters in field space: $f$ and $v \neq v_{\mathrm{EW}}$.}
    \label{fig:ABvev}
\end{figure}

The scale of electroweak symmetry breaking $v_{\mathrm{EW}}$ is defined in terms of the gauge boson masses, as shown in eq.~(\ref{eq:gbmasses}). Since the Higgs particle is among the non-linearly parameterized Goldstone modes, $\sqrt{2} \langle h \rangle = v \neq v_{EW} = 246$~GeV (see figure~\ref{fig:ABvev}). As discussed in ref. \cite{Burdman:2014zta}, this implies that the coupling of the Higgs to the weak bosons would deviate from the SM predicted values by a factor of $\cos \left(\frac{v}{\sqrt{2} f}\right)$.
Exact $\mathbb{Z}_2$ symmetry implies $v_{\textrm{EW}}=f$ and that the Higgs couples with equal strength to both A and B sector gauge bosons. For this reason, exact $\mathbb{Z}_2$ symmetry is not phenomenologically viable.

If we assume that the exact $\mathbb{Z}_2$ is broken such that $v_{\mathrm{EW}} \ll f$, with the partner sector being heavier than the SM, the mass of the Higgs is set roughly as
\begin{equation}
m_h^2 \sim \frac{3 y^2}{8 \pi^2} m^2_{t_B} \log \left( \frac{\Lambda^2}{m^2_{t_B}} \right) \sim {\left( \frac{f}{\pi} \right)}^2,
\label{eq:higgsmass}
\end{equation}
So, for $m_{h} = 125$~GeV, we are led to expect $f \sim 500$~GeV, which also justifies the assumption of $\mathbb{Z}_2$ breaking.
For more details on phenomenological considerations in Twin Higgs models and experimental consequences, see ref.
\cite{Goh:2006wj,Jung:2007ea,Abada:2007sm,Dolle:2007ce,
Batra:2008jy,Liu:2008ca,Liu:2008bx,Liu:2008zzb,
Liu:2009ak,Ma:2009kj,Liu:2009zze,Goh:2009fm,
Shen:2010zzh,Liu:2010jp,Liu:2010nn,
Shen:2011zz,Shen:2011zzg,Zhang:2011as,Wang:2011zzt,Ma:2011zzb,Wang:2011zza,Liu:2011zzi,Wang:2011mra,
Zhan-Ying:2013kxa,Liu:2013dma,Liu:2013gpy,
Han:2014fla,Liu:2014pts,Liu:2014rqa,Burdman:2014zta,Craig:2014roa,Liu:2014uua,
Hetzel:2015bla,Schwaller:2015tja,Han:2015orc,Garcia:2015loa,Craig:2015xla,Garcia:2015toa,Farina:2015uea,Curtin:2015fna}

To achieve a soft breaking of the $\mathbb{Z}_2$ symmetry, a term $\mu^2 H_A^\dagger H_A$ is added to the potential. Note that having a small $\mu^{2}$ is technically natural since it is the only coupling in the theory that violates the $\mathbb{Z}_2$ symmetry. So, for $\mu^{2}\sim m_{h}^{2}$ (which itself arises at one-loop), higher-loop effects of $\mu^{2}$ can be safely neglected.


\section{The Effective Potential at finite temperature}
\label{sec:FTgen}

In this section we review the basic aspects of finite temperature field theory, which we need to compute the effective potential for the Twin Higgs model in the finite temperature equilibrium state. We use the Matsubara formalism for finite temperature calculations \cite{Dolan:1973qd,Kapusta:2006pm}.

Let us consider a renormalizable field theory in the perturbative regime, where we only turn on a background value for one scalar degree of freedom, denoted from here on as $\phi$. Since we are interested in the phase structure of a gauge theory in particular, the scalar in question will be taken to transform nontrivially under a gauge group. The full one-loop finite temperature effective potential for $\phi$ can be split up into a zero temperature part (including one loop effects) and a finite temperature correction
\begin{equation}
V_{\textrm{eff}} (\phi, T) \equiv V_{\textrm{tree}} (\phi) +  V_1^{T=0} (\phi) + \Delta V^T_1 (\phi,T)
\end{equation}
with
\begin{equation}
    \Delta V_1^T (\phi,T) \equiv \frac{T^4}{2 \pi^2} \mathrm{STr} \left[  J_{b/f}  \left( \frac{m_{i}^2 (\phi)}{T^2}\right) \right]
    \label{eq:V1T}
\end{equation}
where for each particle denoted by the label $i$, $m_{i}(\phi)$ denotes its mass in the background $\phi$, and by our assumption of perturbativity $m_{i}(\phi)\lesssim\BigO{\phi}$. The supertrace includes the correct factor accounting for the number of degrees of freedom associated with each particle and a minus sign for fermions. $J_{b}$ and $J_{f}$ arise from the Bose-Einstein and the Fermi-Dirac distribution functions respectively, and they are given as functions of $x_{i} \equiv \frac{m_{i}^2(\phi)}{T^2}$ as
\begin{subequations}
	\begin{align}
    	J_b(x_i) &= \int_0^\infty dt \; t^2 \log \left[ 1 - e^{-\sqrt{x_i + t^2}} \right]
		\\
    	J_f(x_i) &= \int_0^\infty dt \; t^2 \log \left[ 1 + e^{-\sqrt{x_i + t^2}} \right].
	\end{align}
	\label{eq:Jbf}
\end{subequations}
Note that due to the gauge symmetry, any phase of $\phi$ is equivalent, and from this point on we will restrict ourselves to $\phi\ge0$.

While the effective potential is not a gauge invariant object, the value of the potential in the vacuum state is well-defined \cite{Nielsen:1975fs,Patel:2011th,Andreassen:2014eha}. Since we are only interested in the question of whether the symmetry is broken, rather than the details of the phase transition, we can simply investigate whether the global minimum of the finite temperature effective potential occurs at the origin of field space, defined as the point where the gauge bosons are massless\footnote{To be precise, the point where the transverse polarizations of the gauge bosons are massless, at the perturbative level.}.

For any given value of $\phi$, we will mainly be interested in high temperatures $T^2 > \phi^{2}$ which due to perturbativity is equivalent to $T^{2} > m_{i}^2 (\phi)$ as mentioned above, and we will often drop the subscript to write $x<1$ to denote the high temperature regime. In this limit the formulae above can be expanded in an asymptotic series (henceforth referred to as the high-temperature expansion)
\begin{subequations}
	\begin{alignat}{7}
		J_b (x) &= &{}- \frac{\pi^4}{45} &{}+ \frac{\pi^2}{12} x &{}- \frac{\pi x^{\frac{3}{2}}}{6} &{}- \frac{x^2}{32} \log \left( \frac{x}{a_b} \right) &{}+ \ldots & \\
		- J_f (x) &= &{}- \frac{7 \pi^4}{360} &{}+ \frac{\pi^2}{24} x &{} &{}+ \frac{x^2}{32} \log \left( \frac{x}{a_f} \right) &{}+ \ldots &
	\end{alignat}
	\label{eq:HTexpansions}
\end{subequations}
where $a_f= \pi^2 e^{-2 \gamma_E + \frac{3}{2}}$ and $a_b= 16 \pi^2 e^{-2 \gamma_E + \frac{3}{2}}$.

\begin{figure}
        \includegraphics[scale=0.5]{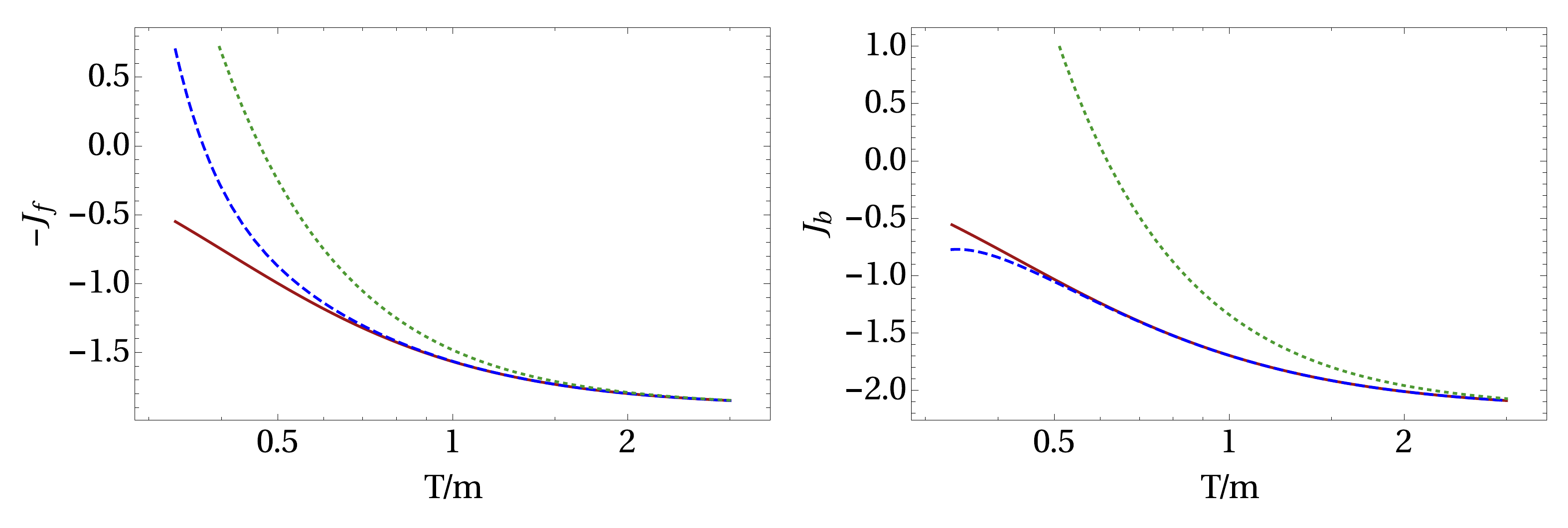}

       \caption{Comparison of different truncations of the high temperature effective potential: fermions to the left and bosons to the right. Solid (red) lines represent the numerical evaluation of eq.~(\ref{eq:Jbf}), dotted (green) lines and dashed (blue) lines respectively represent truncations to linear order in $x$  and a truncation up to and including the logarithmic terms, in eq.~(\ref{eq:HTexpansions}).}
       \label{fig:FTapproximations}
\end{figure}

In figure~\ref{fig:FTapproximations} we compare, for bosons and fermions respectively, a numerical evaluation of equations~(\ref{eq:Jbf}) to the truncation of equations~(\ref{eq:HTexpansions}) at linear order for $x$, and to a truncation up to and including the logarithmic terms. Inspecting the figure, it is evident that the $\BigO{x}$ truncation captures the one-loop effective potential only at very high temperatures ($T\gg m$), while the $\log(x)$ truncation does so at roughly $T \gtrsim{m}$ or even slightly lower temperatures.

Let us consider the salient features of the high-temperature expansion in equation~(\ref{eq:HTexpansions}) term by term, starting with the largest thermal contributions.

{\bf Terms of $\BigO{x^0}$}: Both bosons and fermions have $\phi$-independent Stefan-Boltzmann contributions $\sim T^4$. This does not affect the structure of symmetry breaking.

{\bf Terms of $\BigO{x^1}$}: This is the first order at which $V_{\mathrm{eff}}$ picks up a $\phi$ dependence, and since $T^{4}\,x=T^{2}m_{i}^{2}(\phi) \sim g^{2}T^{2}\phi^{2}$ where $g$ symbolically denotes the strength of coupling between $\phi$ and the particle labeled by the index $i$, the contributions of $\BigO{x}$ provide an effective thermal mass for $\phi$ proportional to $g T$ (masses in the EFT are parametrically smaller than the Matsubara scale $\pi T$). At finite temperature, bosonic and fermionic modes running in a loop contribute to this term with the same sign because they have opposite boundary conditions on the thermal circle. This is connected to the fact that supersymmetry is broken at finite temperature, and
the scalar mass term can acquire large positive corrections $\delta m^{2}_{\textrm{th}} \sim g^2 T^{2}$
in a supersymmetric theory even though contributions to $m^{2}$ cancel at zero temperature.
These contributions to the effective thermal mass of $\phi$ generically drive the scalar background value towards the origin of field space.
For bosons and fermions respectively, one can set up a correspondence between thermal mass corrections and zero temperature divergent mass contributions~\cite{Comelli:1996vm}
\begin{subequations}
	\begin{alignat}{2}
		\textrm{bosons:}& \;\;\; &\frac{\Lambda^2}{16 \pi^2}  &\longrightarrow \frac{T^2}{12}  \\
		\textrm{fermions:}& \;\;\; - &\frac{\Lambda^2}{16 \pi^2}  &\longrightarrow \frac{T^2}{24}.
	\end{alignat}
	\label{eq:LambdaToT}
\end{subequations}

On the other hand, a symmetry that leads to cancellations between the contributions of same-spin particles to the mass of $\phi$ at zero temperature will also induce a cancellation among corresponding thermal mass contributions, which makes symmetry non-restoration at finite temperature a possibility. This is precisely the case in models where the Higgs is embedded as a pNGB in a nonlinear sigma model and the coupling of the Higgs to the heavy fermionic partners arises from higher dimensional terms. This is how the divergences can conspire to cancel at zero temperature, as illustrated in figure~\ref{fig:cancellations}. This cancellation can be preserved when the model is UV-completed into a linear sigma model, which is true in the linear formulation of the Twin Higgs model presented in \cref{sec:THmodel}.

If the $\BigO{x}$ terms can be made to cancel in this fashion, then the phase structure of the model will depend on the effect of the subleading terms in equation~(\ref{eq:HTexpansions}) which therefore must not be neglected. The physics behind these terms is more subtle and we discuss them next.

{\bf Terms of $\BigO{x^{3/2}}$}: In the Matsubara formalism one can expand the fields into their Kaluza-Klein modes around the compact thermal direction. All heavy modes can then be integrated out, leaving us with a dimensionally reduced effective field theory (EFT) of the zero modes, in three (spatial) dimensions. Note that due to their boundary conditions, fermionic degrees of freedom do not have zero modes and therefore the EFT is a theory of scalars and gauge bosons only.
As can be seen in temporal gauge, the gauge boson degrees of freedom arrange themselves into an adjoint scalar $\langle A_\tau \rangle$ and a gauge field $A_i$.
By dimensional analysis in this EFT, corrections to the vacuum energy from zero modes running in loops must be proportional to $m^3 (\phi)$, which is nothing but the $x^{3/2}$ term in equation~(\ref{eq:HTexpansions}).

\begin{figure}
	\begin{center}
	\begin{tabular}{c c c}
		\includegraphics[scale=0.3]{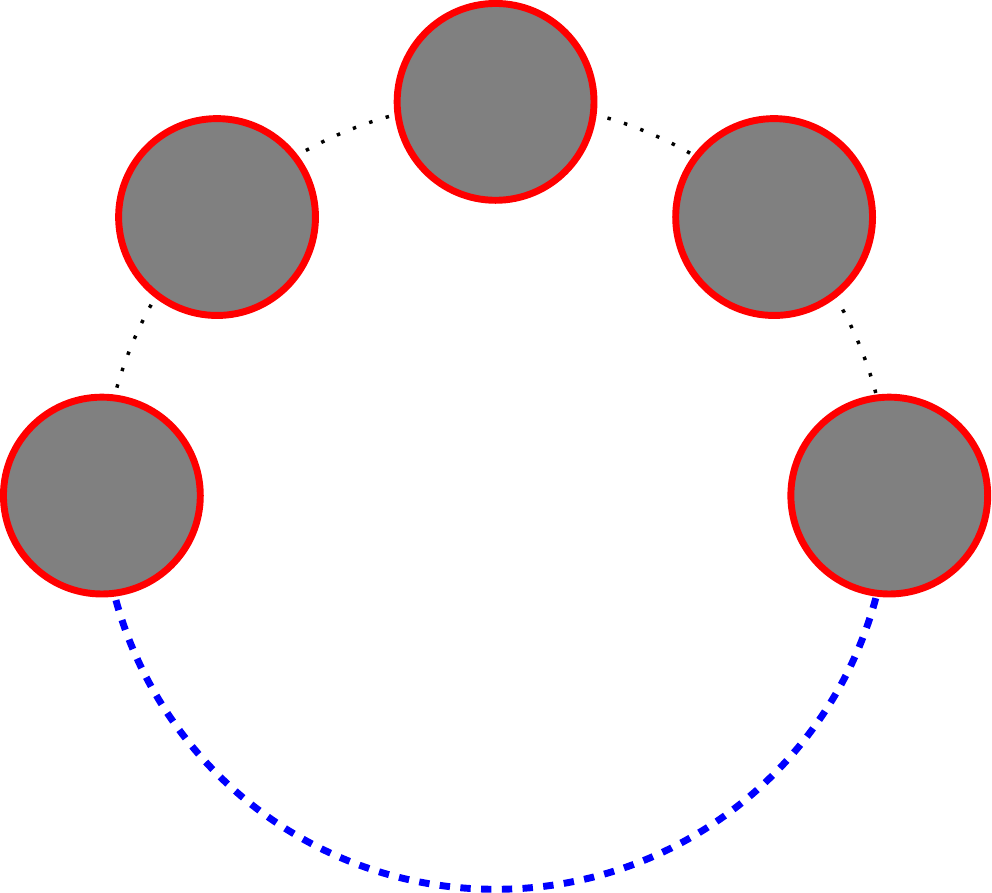}
		&
		$\;\;\;\;\;\;\;$
		&
		\includegraphics[scale=0.7]{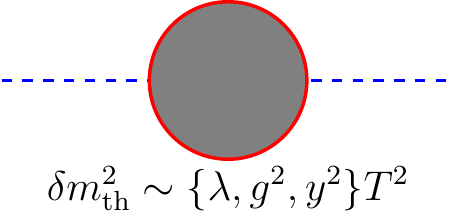}
	\end{tabular}
	\end{center}
	\caption{Schematic diagram of daisy resummation. Blue (dashed) lines correspond to zero-modes and the blobs correspond to the loops involving higher Matsubara modes.}
		\label{fig:daisy}
\end{figure}

Massless zero modes running in loops lead to infrared divergences at higher loop order, which means that a one-loop calculation is inadequate close to the phase transition. A better way to deal with the zero modes is to include the $\BigO{x}$ thermal corrections to their masses discussed above and use the full thermal mass $m^2_{\textrm{th}}(\phi,T)=m^2 (\phi)+\delta m^{2}_{\textrm{th}}(T)$ when calculating their contribution to the one-loop effective potential
\footnote{As shown in ref. \cite{Sakamoto:2007uy} $\langle A_\tau \rangle$ acquires a positive mass, which allows us to restrict our attention to only the 4d scalars as the order parameter for the phase transition.}.
This is equivalent to resumming a series of higher-loop diagrams known as ``ring diagrams'' or ``daisies'' (illustrated in figure~\ref{fig:daisy}) that capture the most egregious infrared divergences, and it is particularly important at temperatures where the thermal mass correction is comparable to, or larger than $m^2(\phi)$.
The ring-corrected finite temperature effective potential thus becomes
\begin{equation}
	V_{\textrm{eff}} (\phi, T) \equiv V_{\textrm{tree}} (\phi) +  V_1^{T=0} (\phi) + \Delta V^T_1 (\phi,T) + \Delta V_{\textrm{ring}} (\phi,T),
\end{equation}
with
\begin{equation}
	\Delta V_{\textrm{ring}} = \sum \; \frac{T^{4}}{12 \pi} \left[
	{\left( \frac{m^2 (\phi)}{T^2}\right)}^{\frac{3}{2}}
	- { \left( \frac{m^2 (\phi) + \delta m^2_{\textrm{th}} (T)}{T^2} \right)}^{\frac{3}{2}} \right].
\end{equation}
where the summation runs over all the scalar degrees of freedom in the dimensionally reduced EFT.

At high enough temperatures $m^2_{\textrm{th}}(\phi,T)$ becomes positive, even for scalars which have $m^{2}(\phi)<0$ at zero temperature. This eliminates contributions to the effective potential coming from the $x^{3/2}$ term that naively appear to be imaginary~\cite{Delaunay:2007wb}. Note that in the Twin Higgs model all scalar modes are pNGBs of the $SU(4)$ symmetry at tree level and therefore do not contribute to the one-loop potential for the SM Higgs which is also among the pNGBs, so this particular issue does not arise.

For zero modes of the transverse polarizations of gauge bosons, residual gauge symmetry in the dimensionally reduced EFT prevents any perturbative mass corrections, including ring diagrams. However, the gauge coupling in the 3d EFT has dimensions of mass, which leads to a non-perturbative mass correction $\delta m_{\mathrm{np}}^{2}\sim g^4 T^2$ \cite{Linde:1980ts}.
Lattice results~\cite{Kajantie:1996mn} indicate that such non-perturbative corrections cannot be neglected in the case of the SM, since they affect the nature of the phase transition and reveal the correct expansion parameter of any perturbative description to be $\frac{m^2_{h}}{m^2_{\textrm{W}}}$.

Even using the best analytical methods available, studying the phase transition is a hard problem~\cite{Morrissey:2012db}. This should not be surprising, since a phase transition corresponds to a non-analyticity in how the free energy depends on the parameters of the model, which cannot be captured at any finite order of perturbation theory \cite{Weinberg:1974hy}. While lattice methods are the most reliable approach in cases such as second order phase transitions, for stronger phase transitions we can gain a qualitative understanding by using analytical methods~\cite{Arnold:1994bp}. We pursue the latter approach here, and hope that our conclusions may serve to motivate further analysis by others.

Taking into account the corrected masses of gauge bosons in the effective theory, their contribution to the effective potential at high temperature has the following form:
\begin{equation} \begin{aligned}
	\sum_{\textrm{polarizations}} T { \left[ m^2(\phi) + \delta m^2_{\textrm{th}} (T) \right] }^{\tfrac{3}{2}} &\approx \sum_{\textrm{polarizations}} \zeta^{\frac{3}{2}} T^4 \left[ 1 + \frac{3}{2} \frac{m^2(\phi)}{\zeta T^2} + \ldots \right] \\
&= \sum_{\textrm{polarizations}} \zeta^{\frac{3}{2}} T^4 + \frac{3}{2} \sqrt{\zeta} T^2 \; m^2 (\phi) + \BigO{T^0}
\end{aligned}\end{equation}
where $\delta m^2_{\textrm{th}} (T) = \zeta T^2$ and $\zeta$ contains numerical factors and couplings. The $T^{4}$ and $T^{2}$ terms imply corrections to the Stefan-Boltzmann term and the thermal mass of $\phi$ respectively, followed by corrections with non-positive powers of $T$. Strictly speaking, at temperatures where non-perturbative thermal mass corrections to the mass of the transverse polarizations of the gauge bosons dominate mass contributions coming from the Higgs mechanism, the Higgs VEV ceases to be a good order parameter.

If a cancellation among the $\BigO{T^2}$ terms in the one-loop effective potential persists in the three dimensional EFT after the resummation, then keeping the subleading terms of $\BigO{\log T}$ becomes crucial to any attempt at an (approximate) analytical study of the phase structure of the theory. For the same reasons as in the discussion of the $\BigO{x}$ terms, this is indeed the case for the model at hand so we finally turn our attention to this last set of terms.

{\bf Terms of $\BigO{\log x}$}:
Note that for each degree of freedom, the logarithmic terms in eq.~(\ref{eq:HTexpansions}) combine with the logarithmic terms in the zero temperature Coleman-Weinberg potential of eq.~(\ref{eq:CW}) to give a $\log \frac{a_{b/f} \; T^2}{\Lambda^2}$ dependence on the temperature, as the factors of $m^{2}(\phi)$ cancel between the one-loop corrections at zero and finite temperature.
Any formal cutoff dependence thus comes from the zero temperature Coleman-Weinberg potential, whose parameters have been chosen to reproduce the observed electroweak VEV and Higgs mass.
As we will see in the next section, the non-cancellation of these terms will determine the fate of electroweak symmetry restoration at finite temperature, in the Twin Higgs model we consider.

Having reviewed the most important aspects of field theory at finite temperature in general, we will apply what we have learned specifically to the Twin Higgs model in the next section.


\section{Twin Higgs at finite temperature}
\label{sec:THfinitetemp}

\begin{figure}[t]
\centering
\includegraphics[scale=0.7]{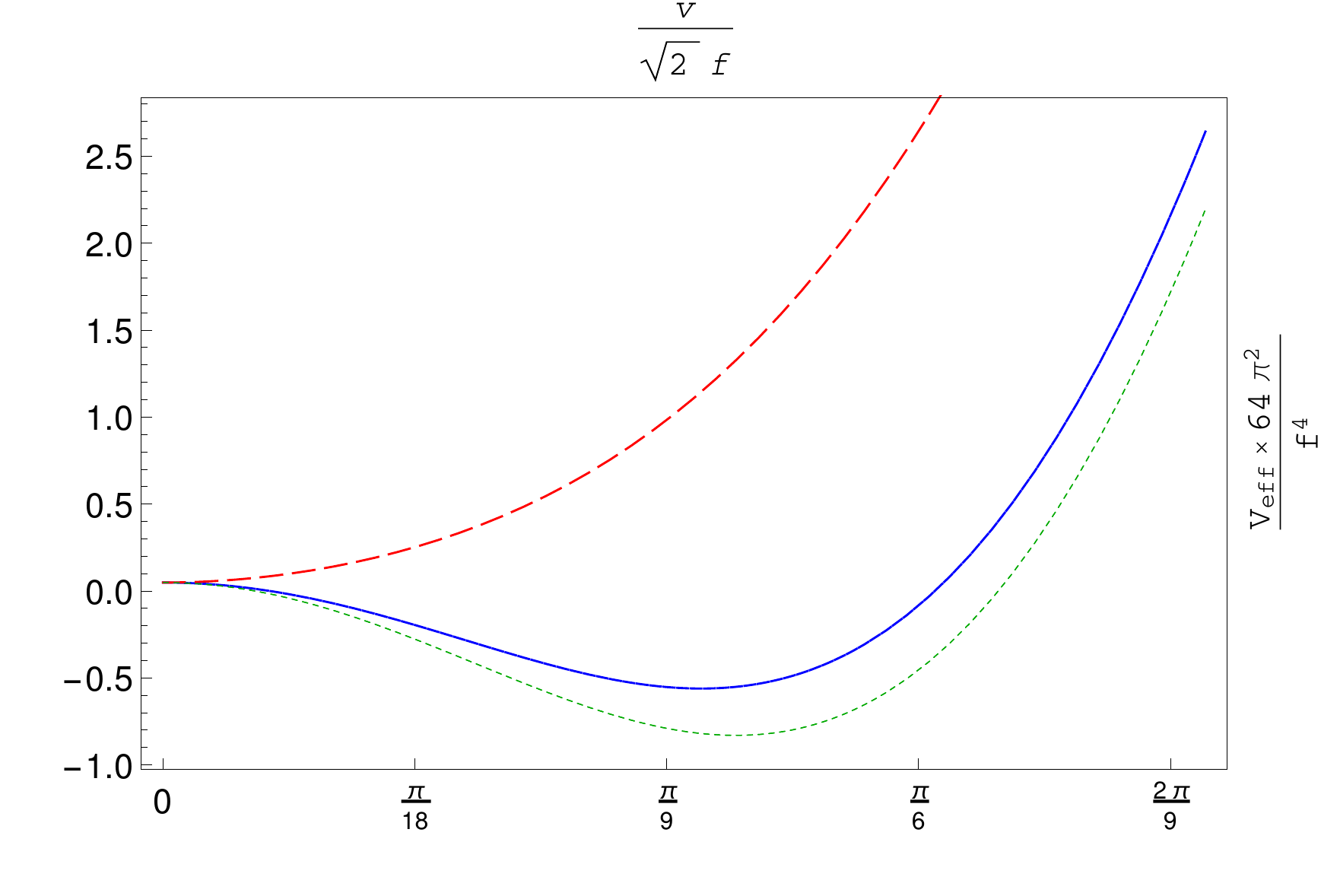}
\caption{We plot the finite temperature effective potential at two different temperatures: The blue solid line represents the potential (numerically evaluated) at $T_1 = 100$~GeV, the red dashed line represents the potential (numerically evaluated) at $T_2 = 350$~GeV and the green dotted line represents the $\BigO{T^2}$ truncation of the potential. Note that for the quadratic truncation, the potential is independent of temperature, and hence, does not sense symmetry restoration. See the main text for the numerical values of the relevant parameters that were used in making the plot.}
\label{fig:NumSymmRest}
\end{figure}

Let us now specialize our discussion to the Twin Higgs model at finite temperature, and let us consider whether there can be any important contributions to the effective potential that we have not already accounted for in the previous section. Due to invariance under gauge symmetries, $H_A^\dagger H_A$ and $H_B^\dagger H_B$ are the only combinations that the effective potential can depend on at zero or finite temperature, and in the exact $\mathbb{Z}_2$ limit these have identical coefficients, which ensures an accidental $SU(4)$ symmetry for quadratic terms, forbidding any dependence on the Goldstone modes at that order. Furthermore, even though the $\mathbb{Z}_{2}$ is broken by the $\mu^{2}$ term introduced at the end of section~\ref{sec:THmodel}, this is a soft breaking, thus any $\mathbb{Z}_{2}$ violating corrections to the potential must include a positive power of $\mu^{2}$. This means that by simple dimensional analysis, at the renormalizable level there can be no $\mathbb{Z}_{2}$ violating contributions to the potential with a positive power of temperature that depend on the Goldstone modes either. There can be contributions of order $\log(T)$ that are $\mathbb{Z}_{2}$ violating, but these are subdominant to the contributions of order $\log(T)$ that have already been considered at the end of section~\ref{sec:FTgen}, since $\mu^{2}\ll f^{2}$.

Of course, the potential has a dependence on the Goldstone modes beyond quadratic order, where $\mathbb{Z}_{2}$ invariance is no longer equivalent to full $SU(4)$ invariance, leading to terms such as $\left(H_A^\dagger H_A\right)^2 + \left(H_B^\dagger H_B \right)^2$. Being dimension four operators, by dimensional analysis the coefficients of such terms also cannot include a positive power of temperature, so these fall into the class of contributions of order $\log(T)$ that we have already discussed.

Since we have now convinced ourselves that all thermal mass corrections of $\BigO{T^2}$ cancel, let us proceed to evaluate the logarithmic contributions from the top quark and its partner:
\begin{equation}
	\begin{aligned}
		- \left( 3 \times 4 \right) \; &\left[\frac{{\left(y^2 f^2 \sin^2 \frac{h}{f} \right)}^2}{64 \pi^2} + \frac{{\left( y^2 f^2 \cos^2 \frac{h}{f} \right)}^2}{64 \pi^2} \right] \log \frac{a_F T^2}{\Lambda^2} \\
		&= - 12 \left[\frac{{\left(y^2 {\left\{h + \ldots \right\}}^2 \right)}^2}{64 \pi^2} + \frac{{\left( y^2 {\left\{ f - \frac{h^\dagger h }{2f} + \ldots \right\}}^2 \right)}^2}{64 \pi^2} \right] \log \frac{a_F  T^2}{\Lambda^2} \\
		&
		\sim\;\; \ldots + \frac{3 y^4 f^2 h^\dagger h}{8 \pi^2} \log \frac{a_F T^2}{ \Lambda^2} + \ldots
	\end{aligned}
	\label{eq:oneloop-thermal-log-mass}
\end{equation}
In hindsight, the fact that these subleading contributions do not cancel each other should come as no surprise. The non-cancellation of logarithmic terms between $\mathbb{Z}_2$ partners in the one-loop Coleman-Weinberg potential is precisely what keeps the Higgs from being an exact Goldstone boson and ensures a sizable Higgs mass at zero temperature (see for instance equation~(\ref{eq:higgsmass})). As mentioned previously, note that the appearance of $\Lambda$ eq.~\ref{eq:oneloop-thermal-log-mass} arises from the zero temperature Coleman-Weinberg potential, and the finite temperature additive corrections are independent of the cutoff (refer eq.~\ref{eq:HTexpansions}).

More generally, in the case of phenomenologically viable SM extensions with a pNGB based mechanism for naturalness, any ultraviolet divergent contribution to the Higgs effective potential at zero temperature will carry over to a corresponding finite temperature contribution.
In the specific case of the Twin Higgs model, as well as in other natural models with a similar symmetry structure, we expect this feature to drive symmetry restoration at finite temperature.

If we wish to study the phase of the theory at temperatures around the electroweak scale (and not significantly higher than the partner masses), it is straightforward to numerically evaluate the one-loop effective potential. While the high-temperature approximations in eq.~(\ref{eq:HTexpansions}) are analytically tractable and help shape our thinking, we choose to numerically evaluate eq.~(\ref{eq:V1T}) and eq.~(\ref{eq:Jbf}) in order to avoid any artifacts from truncating the expansion. The results for our benchmark model (with $f=450$~GeV, $\mu=90$~GeV and $\Lambda = 4.4$~TeV) are presented in figure~\ref{fig:NumSymmRest}. The most important one-loop effects come from the top sector, followed by the electroweak gauge sector, resulting in the restoration of electroweak symmetry at $T \sim 300$~GeV (which stays restored as we push temperatures up to where the EFT starts breaking down).

Since non-perturbative effects cloud the study of physics close to the phase transition, it would be nice to attack this question from a different angle. In particular, it would be of interest to look for a symmetry restored phase at temperatures much higher than the phase transition, where a resummed theory has a valid perturbative description. Of course, this cannot be done in the nonlinear Twin Higgs model and necessitates a UV-completion, which we take to be a linear sigma model completion of the Twin Higgs described in section~\ref{sec:THmodel}. In this UV-completion, the ``radial mode'' linearizes the sigma model. The radial mode is a singlet of the approximate $SU(4)$ global symmetry in the scalar sector, and thus its zero-temperature mass will not be protected from quadratically divergent contributions. This means that at high temperatures, the radial mode picks up a positive thermal mass term, leading to a symmetry restored phase in the UV completion.
The radial mode being driven to zero is a sufficient condition for the gauge bosons to become massless (up to thermal contributions).
Note that this result is not directly related to our calculation in the nonlinear model, as the VEVs of the radial mode and the Goldstone modes are separate from each other.


\section{Conclusions and Outlook}
\label{sec:conc}

We have investigated the possible existence of a broken phase of electroweak symmetry at high temperature in extensions of the Standard Model where the Higgs is realized as a pNGB, focusing on the Twin Higgs model as a benchmark. While we have confirmed that one-loop quadratic contributions to the Higgs potential at finite temperature cancel between the Standard Model degrees of freedom and their partners as they do at zero temperature, this is not true for subleading corrections to the effective potential, which restore electroweak symmetry at high temperature.
Cancellation of $\BigO{T^2}$ corrections to terms in the the Higgs potential is a generic consequence of same-spin partners ensuring naturalness at zero temperature, and the logarithmic corrections are connected to obtaining a phenomenologically viable Higgs boson mass at zero temperature. In the case of the Littlest Higgs model considered in ref.~\cite{Espinosa:2004pn}, the EFT has uncancelled quadratically divergent corrections to higher-order terms in the Higgs potential (arising from non-renormalizable operators), but even in that case, the theory exhibits a restoration of electroweak symmetry as long as temperatures are not pushed beyond the range of validity of the EFT for a finite temperature calculation.

It should be noted that nonlinear sigma models in which the Higgs is a pseudo-Goldstone generically become strongly coupled at high energies and require UV completion, which brings up the question of whether a suitable UV completion may nevertheless allow for a broken phase of electroweak symmetry to persist at high temperature. As we have demonstrated for the case of the Twin Higgs model, UV completing the theory into a linear sigma model cannot achieve this, since the mass of the radial mode is unprotected from quadratic corrections, which at finite temperature drive the radial mode to the origin of field space and lead to symmetry restoration. One can also contemplate nesting one nonlinear sigma model inside another with a higher symmetry breaking scale, however since the Higgs receives a thermal correction (albeit at the subleading level) in the original nonlinear sigma model, this type of construction will not change the finite temperature behavior.

Alternatively, one can imagine supersymmetrizing the linear sigma model, since supersymmetry is the best understood UV-complete mechanism to protect the mass of a scalar from quadratically divergent corrections. However we know that supersymmetry does not prevent quadratic mass corrections at finite temperature, and therefore the radial mode VEV would still be driven to zero.

Let us also briefly remark on classes of natural extensions of the SM other than supersymmetry and Higgs as a pNGB. Theories with strongly coupled Higgs sectors appear to be disfavored in light of the experimental findings at the LHC, and in any case these typically exhibit symmetry restoration for temperatures above the formation of the condensate. Gauge-Higgs models have been shown to lead to a restored symmetry phase at high temperature~\cite{Panico:2005ft}. In ``Relaxion'' models~\cite{Graham:2015cka}, electroweak symmetry is also restored at high temperatures, but this idea is quite recent and it would be interesting to study whether variants of it may have a more subtle finite temperature behavior.

Our conclusions are also consistent with more general considerations based on the thermodynamic behavior of systems at high temperature. In particular, the free energy of a system is given by
\begin{equation}
\mathcal{F} = \mathcal{E} - T \mathcal{S},
\end{equation}
and therefore, heuristically, at high temperatures, the free energy can be minimized by increasing entropy (corresponding to a symmetric phase) rather than lowering the energy by spontaneous symmetry breakdown~\cite{Orloff:1996yn}. This suggests a robust rule of thumb that symmetries get restored at high temperatures, in the absence of any other thermodynamic variables describing the system that can attain values that are ``natural'' based on dimensional analysis. We remark in passing that if this last criterion is removed, e.g. when the system has a chemical potential $\mu\sim T$, symmetry non-restoration is possible, see for example refs.~\cite{Linde:1976kh,Haber:1981ts,Benson:1991nj,Riotto:1997tf}.


\section{Acknowledgements}
The authors would like to thank Zackaria Chacko, Nathaniel Craig, David Morrissey, Raman Sundrum and Jiang-Hao Yu for helpful discussion and valuable comments. CK would also like to thank Adam Falkowski, Rakhi Mahbubani, Georgios Pastras and Devin Walker, with whom he has studied related questions in the past. The research of the authors is supported by the National Science Foundation under Grants No. PHY-1315983 and No. PHY-1316033.

\bibliographystyle{JHEP}
\bibliography{references}

\end{document}